\documentclass[aps,prb,showpacs,reprint,amsmath,amssymb,superscriptaddress,a4paper]{revtex4-1}
\usepackage[utf8]{inputenc}
\usepackage{amsmath}
\usepackage{amsfonts}
\usepackage{amssymb}
\usepackage{graphicx}
\usepackage{lineno}

\begin{document}

\title{Shadow epitaxy for \emph{in-situ} growth of generic semiconductor/superconductor hybrids}

\author{Damon J. Carrad}
\thanks{These two authors contributed equally}
\affiliation{Center for Quantum Devices, Niels Bohr Institute, University of Copenhagen, 2100 Copenhagen, Denmark.}

\author{Martin Bjergfelt}
\thanks{These two authors contributed equally}
\affiliation{Center for Quantum Devices, Niels Bohr Institute, University of Copenhagen, 2100 Copenhagen, Denmark.}

\author{Thomas Kanne}
\affiliation{Center for Quantum Devices, Niels Bohr Institute, University of Copenhagen, 2100 Copenhagen, Denmark.}

\author{Martin Aagesen}
\affiliation{Center for Quantum Devices, Niels Bohr Institute, University of Copenhagen, 2100 Copenhagen, Denmark.}
\affiliation{Danish Defence Research Center, 2750 Ballerup, Denmark.}

\author{Filip Krizek}
\affiliation{Center for Quantum Devices, Niels Bohr Institute, University of Copenhagen, 2100 Copenhagen, Denmark.}
\affiliation{Department of Spintronics, Institute of Physics, Czech Academy of Sciences, 162 00 Praha 6, Czech Republic.}

\author{Elisabetta M. Fiordaliso}
\affiliation{DTU Nanolab, Technical University of Denmark, 2800 Kgs. Lyngby, Denmark.}

\author{Erik Johnson}
\affiliation{Center for Quantum Devices, Niels Bohr Institute, University of Copenhagen, 2100 Copenhagen, Denmark.}
\affiliation{Department of Mechanical Engineering, Technical University of Denmark, 2800 Kgs. Lyngby, Denmark.}

\author{Jesper Nyg{\aa}rd}
\email{nygard@nbi.ku.dk}
\affiliation{Center for Quantum Devices, Niels Bohr Institute, University of Copenhagen, 2100 Copenhagen, Denmark.}

\author{Thomas Sand Jespersen}
\email{tsand@nbi.ku.dk}
\affiliation{Center for Quantum Devices, Niels Bohr Institute, University of Copenhagen, 2100 Copenhagen, Denmark.}


\begin{abstract} Uniform, defect-free crystal interfaces and surfaces are crucial ingredients for realizing high-performance nanoscale devices. A pertinent example is that advances in gate-tunable and topological superconductivity using semiconductor/superconductor electronic devices are currently built on the hard proximity-induced superconducting gap obtained from epitaxial indium arsenide/aluminium heterostructures. Fabrication of devices requires selective etch processes; these exist only for InAs/Al hybrids, precluding the use of other, potentially superior material combinations. We present a crystal growth platform -- based on three-dimensional structuring of growth substrates -- which enables synthesis of semiconductor nanowire hybrids with \emph{in-situ} patterned superconductor shells. This platform eliminates the need for etching, thereby enabling full freedom in choice of hybrid constituents. We realise and characterise all the most frequently used architectures in superconducting hybrid devices, finding increased yield and electrostatic stability compared to etched devices, along with evidence of ballistic superconductivity. In addition to aluminium, we present hybrid devices based on tantalum, niobium and vanadium.

This is the submitted version of the manuscript. The accepted, peer reviewed version is available from Advanced Materials: http://doi.org/10.1002/adma.201908411

\end{abstract}

\maketitle


One-dimensional semiconductor (SE) nanowires (NWs) proximity coupled to superconductors (SU) have attracted considerable attention from the condensed matter community since the prediction\cite{LutchynPRL10,OregPRL10} and observation\cite{MourikScience12,DasNatPhys12,DengScience16} of Majorana zero-modes, which have been proposed as a basis for topologically protected quantum information processors.\cite{SarmaPRL05,AasenPRX16} To ensure topological protection, methods for growing disorder-free `hard-gap’ SE/SU epitaxial hybrids were developed.\cite{KrogstrupNatMat15,ChangNatNano15,KangNL17} These materials use bottom-up crystal growth of InAs NWs with uniform epitaxial aluminium coatings, an approach which has been extended to high mobility two-dimensional systems\cite{KjaergaardNatComm16, DrachmannNL17} and selective area grown networks.\cite{KrizekPRM18, VaitiekenasPRL18sag} The success of epitaxial InAs/Al hybrids lies in the ability to selectively remove the Al \emph{via} top-down processing, and thereby realise important device classes such as normal metal spectroscopic devices,\cite{DengScience16,ChangNatNano15,KjaergaardNatComm16,DrachmannNL17} Josephson Junctions\cite{vanWoerkomNatPhys17,LarocheNatComm19,TosiPRX19,HaysPRL18} for gate-controlled transmon qubits,\cite{LarsenPRL15, LuthiPRL18} and superconducting Majorana islands.\cite{AlbrechtNature16, ShermanNatNano16,ProutskiPRB19} A limitation of this method is that the need for post-process etching inherently limits materials choice. For instance, despite strong incentives to utilise technologically important superconductors such as Nb~\cite{GuskinNanoscale17} and NbTiN~\cite{GulNL17} -- which exhibit higher transition temperatures, critical magnetic fields and superconducting energy gaps -- selectively removing Nb from InAs remains an unsolved problem. Similarly, while InSb is an attractive semiconductor due to its high mobility, $g$-factor and strong spin-orbit coupling,\cite{GulNL17,ZhangNatComm17,GulNatNano18,ZhangNature18} there is no reported method for selectively removing even Al from InSb without damage. Thus, most potential improvements in epitaxial SE/SU technology are predicated on developing a materials-independent method for device fabrication. An attractive approach to eliminating etch processes is to employ an \emph{in-situ} `shadow approach' that masks specific segments along the NW from superconductor growth. Initial progress was demonstrated in Refs. \citenum{KrizekNL17,GazibegovicNature17} where deterministically positioned NWs were shadowed by adjacent NWs. This approach, however, requires accurate control of relative NW positions and growth directions. Further, the range of the possible device geometries is limited, since the NWs create only narrow gaps in the SU coating.

\begin{figure*}[t]
\includegraphics[width=17cm]{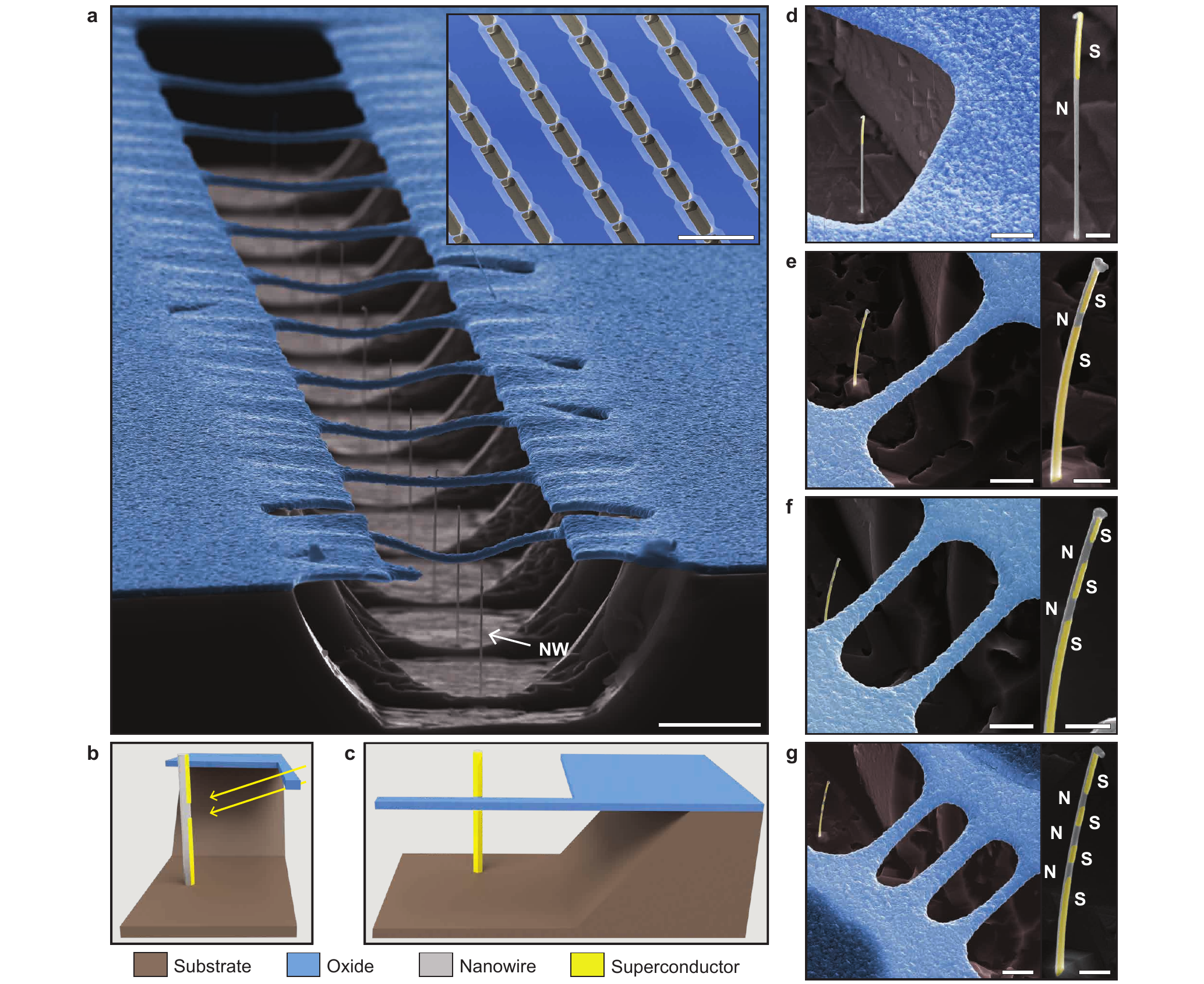}
\caption{\textbf{Shadow epitaxy platform for hybrid nanowire growth. a}, False colour scanning electron micrograph (SEM) of SiOx (blue) bridges spanning trenches in an InAs (grey) substrate. InAs nanowires (NW) are grown in proximity to the bridges, which act as a shadow mask. Inset: Overview SEM of bridge substrate before growth. \textbf{b,c}, Schematics viewed from \textbf{b}, side and \textbf{c}, the direction of superconductor (yellow) deposition. Direction of superconductor deposition is shown by the arrows in \textbf{b}. The bridge geometry is projected onto each NW to lithographically define the regions left uncoated during superconductor deposition. \textbf{d-h}, False colour SEMs of as-grown NW hybrids with Al overlayer (yellow) with \textbf{d}, half-shadowed/tunnel spectroscopy, \textbf{e}, Josephson junction, \textbf{f}, island and \textbf{g}, double island geometries. S and N indicate the superconducting and bare NW segments, respectively. Scale bars represent \textbf{a}, $5~\mu$m (main) and $50~\mu$m (inset) \textbf{d-g}, $2~\mu$m (left images) and $500$~nm (right images).} 
\end{figure*}

\begin{figure}[t]
\includegraphics[width=8.5cm]{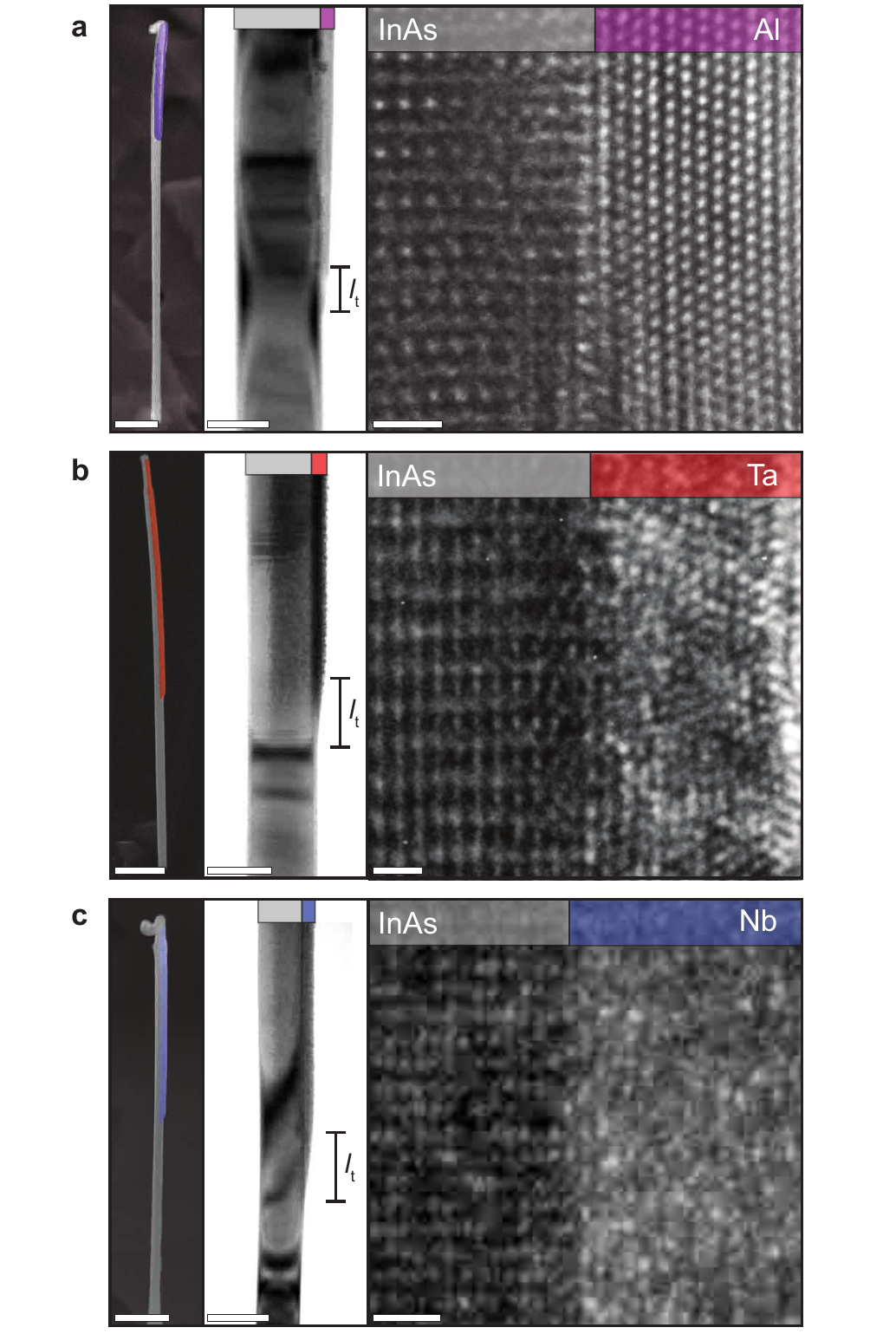}
\caption{\textbf{SEM (left panels) and TEM (middle and rightmost panels) of half-shadowed, half-shell hybrid materials grown using the shadow epitaxy platform. a}, Al/InAs shadow hybrid with 8~nm thick Al deposited at $T = -150^\circ$C, exhibiting an epitaxially matched interface.\cite{KrogstrupNatMat15} \textbf{b}, Tantalum (middle panel 20~nm thick; right panel 5~nm thick) and \textbf{c}, niobium (40~nm thick) films exhibited nanocrystalline/amorphous structure. The length $l_{\mathrm{t}}$ of the tail at the SU shadow edge was determined by the geometry of the mask and chamber, as well as material diffusion (Supplementary Section 2). Ta and Nb had typical $l_{\mathrm{t}}~\sim~65-120$~nm and Al $l_{\mathrm{t}} \leq 60$~nm. The broad bright/dark fringes in low-resoultion TEM (middle panels) are likely bending contours, whereby slight bends in the NW result in modulation of the crystal plane orientation with respect to the beam. Scale bars represent 500~nm (left images), 100~nm (middle images) and 1~nm (right images).} 
\end{figure}

\begin{figure*}[t]
\includegraphics[width=17cm]{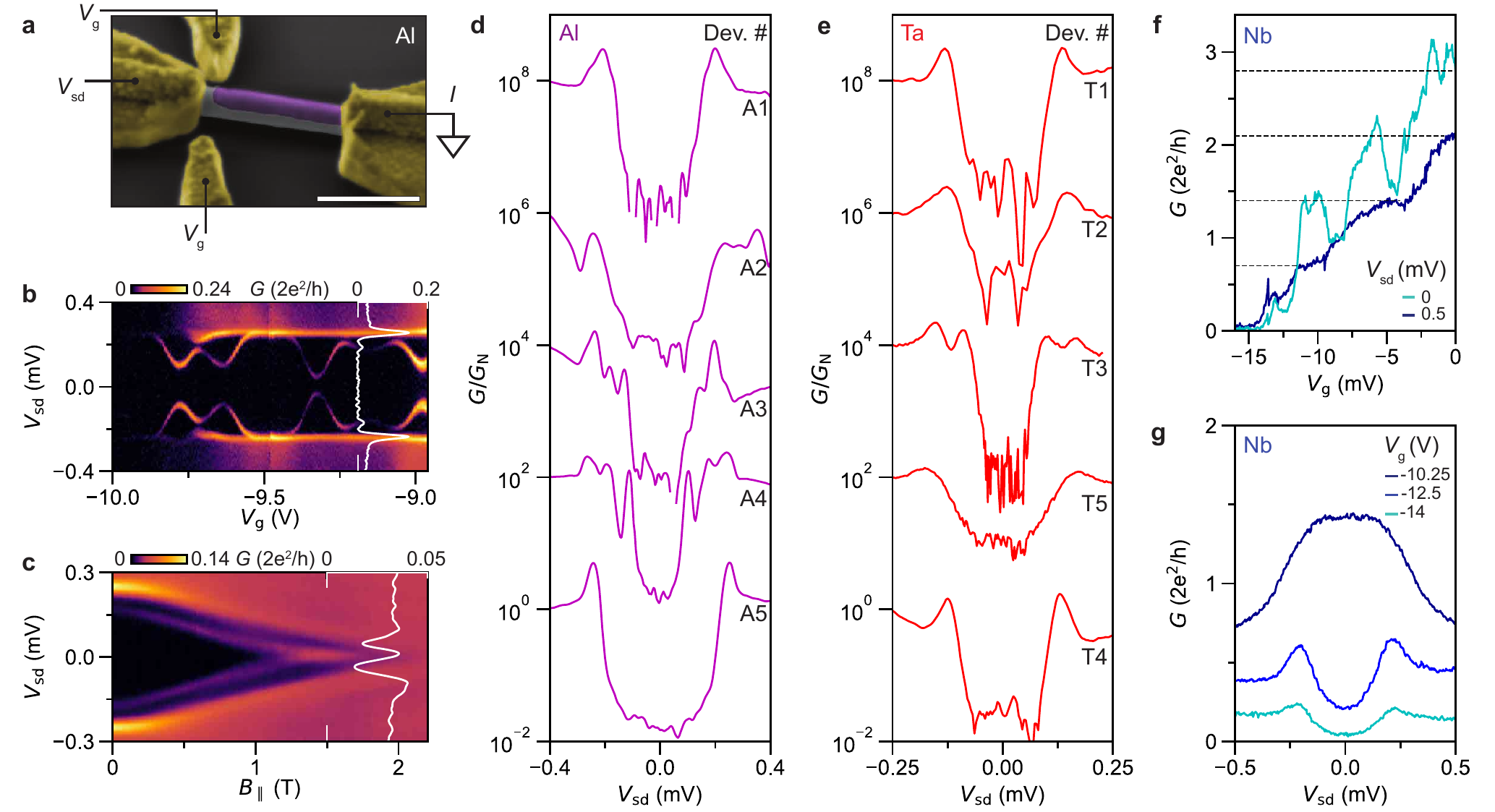}
\caption{\textbf{Induced superconductivity in Al/InAs, Ta/InAs and Nb/InAs hybrids. a}, False-color SEM of InAs NW (grey) with shadow patterned Al (purple) and \emph{ex-situ} Ti/Au contacts (gold). Scale bar represents 500~nm. \textbf{b}, Differential conductance $G~=~dI/dV_{sd}$ vs. $V_{\mathrm{sd}}$ and $V_{\mathrm{g}}$ close to pinch-off. A continuous, hard gap with $\Delta=0.25$~meV and sub-gap states are evident. \textbf{c}, Measured $G$ vs.\ $B_{\parallel}$ for $V_{\mathrm{g}}$ fixed at the position of the line trace in $\textbf{b}$ shows the gap closing at $B_{\mathrm{C}}~\sim~2$~T. A peak at zero bias emerges at $B=1.3$~T from coalescing bound states.\cite{DengScience16} \textbf{d}, Typical measurements of $G$ (normalised to $G_{\mathrm{N}}$ at $V_{\mathrm{sd}}=-0.4$~mV) vs.\ $V_{\mathrm{sd}}$ on a logarithmic scale. A hard superconducting gap is found for all five Al/InAs devices, A1-A5. \textbf{e}, Corresponding measurement for Ta/InAs devices ($G_{\mathrm{N}} = G(V_{\mathrm{sd}}=-0.25$~mV)). Traces in \textbf{d,e} are offset by two decades for clarity. \textbf{f}, $G$ vs.\ $V_g$ for a Nb/InAs device exhibiting conductance plateaus for $V_{\mathrm{sd}}=0.5$~mV~$>~\Delta$ at multiples of $0.7~\times~2e^2/h$ (dashed lines), and conductance doubling in the superconducting state, $V_{\mathrm{sd}}=0$. \textbf{g}, $G$ vs $V_{\mathrm{sd}}$ shows conductance doubling on the first plateau at $V_{\mathrm{g}}=-10.25$~V, and a soft gap in the tunnelling regime, $V_{\mathrm{g}}=-12.5, -14$~V. Contact resistance $R_\mathrm{C} = 3.2$~k$\Omega$ was subtracted from the data in \textbf{f,g.}}
\end{figure*}

Here, we demonstrate a flexible platform for growing device-ready hybrids in numerous geometries. We combine pre-growth fabrication of shadow structures directly on NW growth substrates with precise positioning of NW growth sites and full \emph{in-situ} control of the orientation of superconductor flux. The shadow epitaxy platform enables independent materials choice (SU,SE) and we demonstrate simultaneous growth of the most important hybrid device architectures using aluminium, niobium, tantalum and vanadium.\cite{BjergfeltNanotech19} Eliminating post-process chemical etching increased spectroscopic device yield (5 out of 5 for Al and Ta devices), enabled ballistic superconductivity in Nb devices, and enhanced electrostatic stability by an order of magnitude in Majorana island devices. Very high yield and large-range, hysteresis-free device operation are crucial metrics for future advanced circuitry that incorporates a large number of hybrid elements. Finally, we show that the platform is compatible with both half-shell and full-shell hybrid geometries and allows for functional devices to be encapsulated \emph{in-situ} with passivating dielectrics, thus protecting sensitive elements. While the focus here is on semiconductor/superconductor hybrids, shadow epitaxy is equally applicable for \emph{any} metal/semiconductor/insulator combination, extending the scope to other applications where pristine surfaces and interfaces are key.\cite{delAlamoNat11, RielMRS14}

\section*{The shadow epitaxy platform}
\textbf{Figure~1} illustrates the platform features. A (111)B-oriented InAs growth substrate is patterned with etched trenches and a series of silicon oxide (SiOx) `bridges'. NWs are grown from gold catalyst particles, pre-positioned at the trench bottom with the desired lateral distance from the overhanging bridge(s). NWs are grown by molecular beam epitaxy\cite{KrogstrupNatMat15} and the bridges act as shadow masks for subsequent \emph{in-situ} deposition of superconductors, normal metals, or dielectrics. The material is deposited from a direction parallel to the trenches and inclined by an angle $\theta = 5-45^{\circ}$ from the substrate surface. The situation is schematically shown in Figure~1b,c. The bridge design is thereby projected as a pattern in the superconducting layer on each NW, effectively growing the desired device architecture, with independent choice of SE and SU material. For process details see Methods and Supplementary Section~1.

To ensure a pristine interface, the sample remains under ultra-high vacuum ($< 10^{-8}$~Torr) between NW growth and superconductor deposition.\cite{KrogstrupNatMat15, BjergfeltNanotech19, GuskinNanoscale17,KangNL17,KjaergaardNatComm16} In the simplest, single-deposition case, the superconductor coats 2 or 3 of the NW's 6 facets -- i.e. a half-shell coating -- except for regions shadowed by the bridge. This breaks the NW into a sequence of SU segments separated by bare SE. The length and number of segments are controlled by $\theta$ and the projected bridge design (Supplementary Section~2). Additional possibilities using multiple depositions, angles and/or materials are demonstrated below (Figure~5). Figure~1e-g show scanning electron micrographs (SEMs) of bridges designed to produce the four most important hybrid device geometries. Half-shadowed NWs (Figure~1d) are the canonical design for tunnel-spectroscopy characterisation of the (sub)gap properties of hybrids,\cite{ChangNatNano15, MourikScience12, DengScience16, ZhangNature18} and are obtained by shadowing the entire lower section of the NW with a wide bridge. A single, narrow bridge (Figure~1e) produces a gate-tunable Josephson junction,\cite{DohScience05} the component at the heart of gatemon and Andreev qubits.\cite{LarsenPRL15, LuthiPRL18, vanWoerkomNatPhys17,LarocheNatComm19,TosiPRX19,HaysPRL18} Double and triple bridges (Figure~1f,g) produce hybrids with single\cite{AlbrechtNature16,HigginbothamNatPhys15,ShenNatComm18} and double\cite{ShermanNatNano16} Majorana island geometries, respectively, which are the building blocks of topological quantum computation schemes.\cite{AasenPRX16, PluggeNJP17} In summary, the shadow epitaxy platform enables materials-independent, wafer-scale, parallel synthesis of device-ready hybrids, in many different geometries.

\section*{Patterning and structural characterisation of SU films}
The SU and SE feature sizes are largely defined by the shadow epitaxy mask used, although evaporation chamber geometry and material diffusion\cite{KrogstrupNatMat15} become important when high precision is required. In particular, the finite evaporation source size and the distances between NW, mask and source generate `tail regions' between fully covered and fully shadowed segments (Supplementary Section 2). The expected tail length is $l_\mathrm{t} = 100-200$~nm for our mask/chamber dimensions. To characterise this, \textbf{Figure~2} shows SEM and transmission electron micrographs (TEMs) of InAs hybrids with Al, Ta and Nb shells. Depositions were performed using e-beam evaporation at room temperature except for Al, where $T \sim -150^\circ$C (see Methods), with thicknesses $t_\mathrm{Al} = 8$~nm, $t_\mathrm{Nb} = 40$~nm, $t_\mathrm{Ta} = 20$~nm (left and middle panel) and $t_\mathrm{Ta} = 5$~nm (right panel). Figure~2 middle panels show typical $l_\mathrm{t} = 65-120$~nm for Nb/Ta and $l_t < 60$~nm for Al. The smaller-than-estimated $l_\mathrm{t}$ suggests that material diffusion acts to shorten $l_\mathrm{t}$. The enhanced diffusion and associated shorter $l_\mathrm{t}$ for Al is expected from the lower heat of vaporisation; $\Delta H_\mathrm{vap}^\mathrm{Al} \sim 300$~kJ/mol, c.f. $\Delta H_\mathrm{vap}^\mathrm{Ta,Nb} \sim 700$~kJ/mol.\cite{KrogstrupNatMat15,BjergfeltNanotech19} Note that these effects can be compensated for by, e.g., altering mask design and NW position, using a smaller deposition source or collimator, and/or double-angle evaporation (Supplementary Section~2).

The semiconductor/superconductor interface properties crucially influence the quality of the induced superconducting gap\cite{KrogstrupNatMat15,ChangNatNano15} and the rightmost panels in Figure~2 show high resolution TEM of each interface. All interfaces were uniform, with clear semiconductor termination and no interface oxides/contamination. Al and InAs exhibited an epitaxial relation, with the Al growing with $\langle 111 \rangle$ out-of-plane orientation. This confirms that shadowed interfaces and e-beam deposited Al retain the high quality and structure of MBE-grown hybrids.\cite{KrogstrupNatMat15, GuskinNanoscale17} The Ta and Nb films were amorphous/nanocrystalline with a columnar morphology related the deposition angle, as previously observed for Nb~\cite{GuskinNanoscale17} (Supplementary Section 3). Reliable characterisation of nanocrystalline/amorphous films is challenging due to overlapping signals from randomly oriented grains. However, by imaging a thin ($5$~nm) Ta film, randomly oriented grains with diameter $2-5$~nm can be resolved. An amorphous/nanocrystalline structure was found for Nb depositions both with substrate temperature $T \sim -150^\circ$C and at room temperature.

\section*{Tunnel spectroscopy devices}
Low temperature transport experiments were performed both to benchmark the shadow-patterned hybrids against the wide literature on etched epitaxial Al\cite{KrogstrupNatMat15, ChangNatNano15, DengScience16, AlbrechtNature16, HigginbothamNatPhys15, ShermanNatNano16, ShenNatComm18, VaitiekenasPRL18sag, VaitiekenasPRL18gfactor,KangNL17} -- performance gains are expected due to eliminating post-process etching (Supplementary Section 4) -- and to characterise the new Ta and Nb-based hybrids. \textbf{Figure~3} displays results of tunnel-spectroscopy measurements utilising half-shadowed hybrids (Figure~1d). The exposed SE segment was contacted by a normal metal (Ti/Au) and electrostatic gates were used to induce a quantum dot (QD) or quantum point contact (QPC) tunnel barrier between the Ti/Au and proximitised region.\cite{MourikScience12,LeeNatNano14,ChangNatNano15,DengScience16,KjaergaardNatComm16} The differential conductance, $G = dI/dV_{\mathrm{sd}}$, in the tunneling regime is proportional to the hybrid density of states, as shown in Figure~3b for Al/InAs. A hard superconducting gap is seen for all $V_\mathrm{g}$ with coherence peaks at $V_{\mathrm{sd}} = \pm \Delta/e = \pm 0.25$~mV, highlighted by the line trace at $V_\mathrm{g} = -9.195$~V. The gap hardness $G_{\mathrm{N}}/G_{\mathrm{S}}$ (where $G_{\mathrm{S}} = G(V_{\mathrm{sd}}=0$~mV) and $G_{\mathrm{N}}$ is the out-of-gap conductance $G(V_{\mathrm{sd}}=-0.4$~mV)) was $\sim~75$ throughout the studied $V_\mathrm{g}$ range, with a peak value of 350 (Supplementary Section 5). This is the highest reported figure in a SE/SU hybrid, confirming the high quality of the interface. Figure~3c shows the evolution in parallel magnetic field, $B_{\parallel}$, with $V_{\mathrm{g}}$ fixed at the position of the line trace of Figure~3b. The gap closes at critical field $B_{\mathrm{C}} \sim 2$~T; consistent with the thin ($8$~nm), flat, epitaxial film.~\cite{DengScience16} Coalescing bound states that stick to zero energy for $B_{\parallel} > 1.3$~T were also observed -- highlighted by the line trace taken with $B_{\parallel} = 1.5$~T -- resembling previously reported topological zero modes.\cite{MourikScience12,DengScience16} While a detailed analysis of zero-modes is outside the scope of this article, the magnetic field-compatibility and effective $g$-factor = 6.5\cite{VaitiekenasPRL18gfactor} highlight the potential of the shadow hybrids for studying the topological regime. 

In total, we fabricated five Al/InAs tunnel devices, A1-A5, and logarithmic line traces of normalised conductance $G/G_{\mathrm{N}}$ vs $V_{\mathrm{sd}}$ at fixed $V_{\mathrm{g}}$ are shown in Figure~3d. All five devices exhibited similar behaviour and a hard superconducting gap, with $G_{\mathrm{N}}/G_{\mathrm{S}} = 50 - 100$. Devices A2-A4 feature multiple peaks below the coherence peaks at $\Delta$ due to bound states (Supplementary Section 6). Overall, the results confirm the high quality of the Al/InAs hybrids synthesised using shadow epitaxy, and the device yield (5 out of 5) compares favourably with conventional device processing using wet etching (Supplementary Section 4).

Turning to devices grown with Ta and Nb, all five Ta/InAs devices exhibited induced superconductivity (Fig~3e). Devices T1-T4 showed hard gaps with values $G_{\mathrm{N}}/G_{\mathrm{S}}~=~50$--100, persisting over a wide $V_\mathrm{g}$ range (Supplementary Section 6), similar to Al/InAs hybrids. T5 showed a lower hardness $G_{\mathrm{N}}/G_{\mathrm{S}}\sim~10$. Since the Ta is nanocrystalline/amorphous (Figure~2b) these results show that the atomic ordering of epitaxial interfaces is not a prerequisite for hard gap superconductivity. Rather, an impurity-free, uniform interface is sufficient. The values for $\Delta^{\mathrm{Ta}} = 0.13$~meV and corresponding $T_{\mathrm{C}}^{\mathrm{Ta}} = 0.7$~K are consistent with bulk films of matching thickness, 20~nm (Supplementary Section 7).\cite{HauserRevModPhys64} The Ta/InAs out-of-plane critical field $B_{\mathrm{C}\perp}^{\mathrm{Ta}}~\sim~3.5$~T (Supplementary Section 7) is significantly higher than for Al/InAs $B_{\mathrm{C}\perp}^{\mathrm{Al}}~\sim~100$~mT with similar dimensions, making Ta/InAs hybrids potentially attractive for studies of topological superconductivity.

The technological importance of niobium based-superconductors -- owing to their high $T_\mathrm{C} \sim 9$~K and high critical magnetic fields -- has motivated efforts to incorporate them in \emph{ex-situ}-fabricated hybrid nano-devices.\cite{MourikScience12,GulNL17,ZhangNatComm17,GharaviNanotech17,DrachmannNL17} Shadow epitaxy enables \emph{in-situ} Nb/InAs devices previously impossible due to lack of selective process techniques. Figures~3f,g present low-temperature spectroscopic results. Upon increasing $V_\mathrm{g}$, the conductance for $V_{\mathrm{sd}} = 0.5$~mV~$>~\Delta$ increases in steps of $0.7 \times 2$e$^2$/h and shows a doubling at the plateaus for $V_{\mathrm{sd}} = 0$. This is consistent with a near ballistic junction and near perfect transmission at the contacts.\cite{ChuangNL13,KjaergaardNatComm16, ZhangNatComm17, GulNatNano18} In the tunnelling regime, $V_{\mathrm{g}} \leq -12.5$~V, conductance suppression is clearly observed for $eV_\mathrm{sd} < |\Delta| \sim 0.2$~meV in Figure\ 3g, albeit with relatively low hardness ($G_\mathrm{N}/G_\mathrm{S} \sim 3$). Considering the near ballistic device characteristic and high contact transmission, and given the excellent gap hardness achieved in Al/InAs and Ta/InAs shadow devices, it is unlikely the soft superconducting gap for our \emph{in-situ} Nb/InAs hybrid was caused by contaminated or process-damaged interfaces. Rather, native oxides at the Nb surface could provide an explanation. Nb oxides can be superconducting with lower $T_{\mathrm{C}} \sim 2$~K, metallic, or magnetic~\cite{ProslierAPL08, HalbritterAPA87}; each of these effects leads to soft-gap superconductivity in thin niobium films. Using \emph{in-situ} surface passivation of Nb hybrids would likely provide insights into how the surface impacts gap properties (see Figure~5a). To summarise, the main result of Figure~3 is the demonstration of high-quality hybrid devices with expanded choice of SU and SE.

\begin{figure*}[t]
\includegraphics[width=17cm]{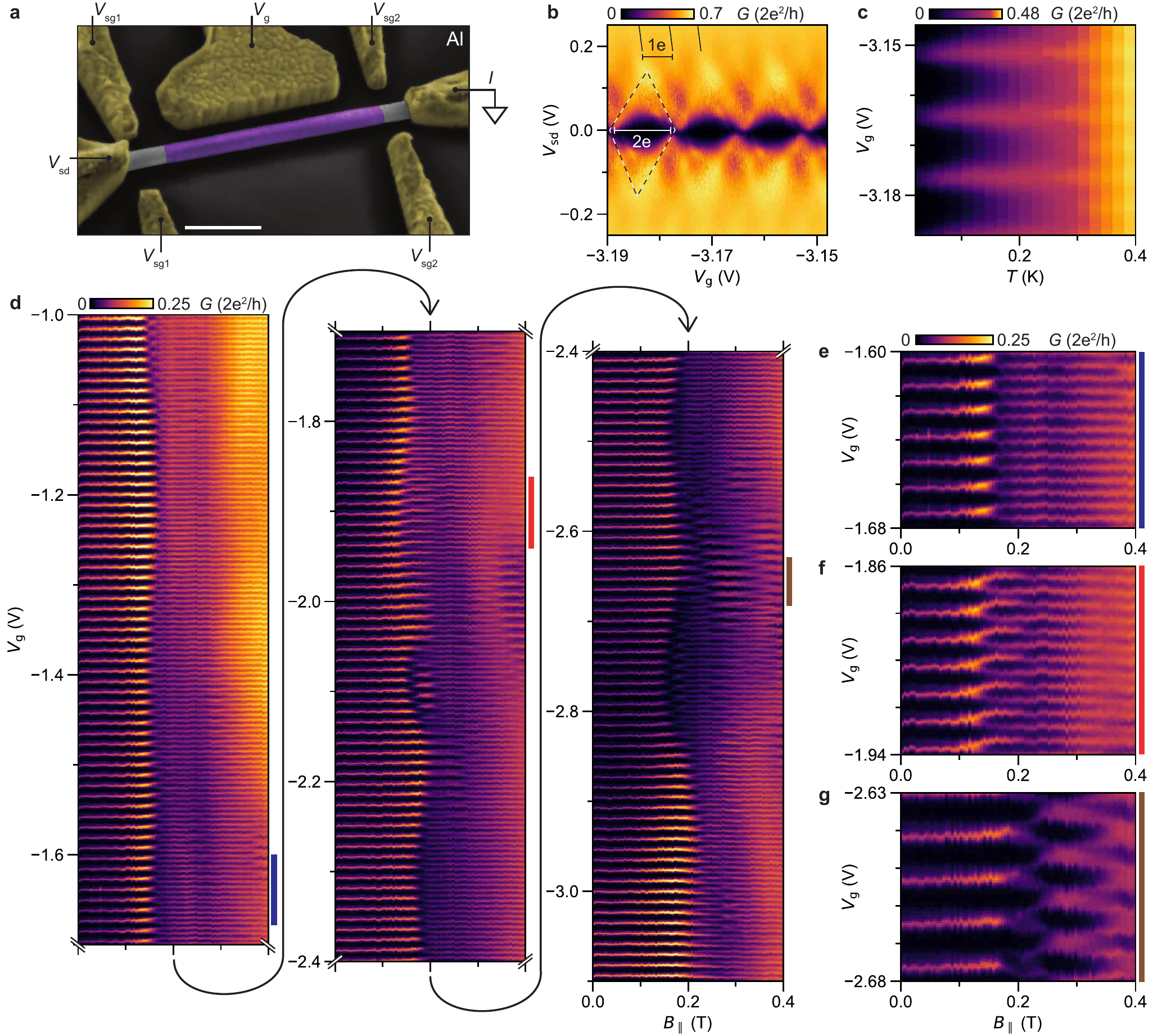}
\caption{\textbf{Electron transport in a Majorana island. a}, False color SEM of shadow patterned Al/InAs island device. Middle gate voltage, $V_{\mathrm{g}}$, was varied, with the four other gates used to generate tunnel barriers. Scale bar represents 500~nm. \textbf{b}, Bias spectroscopy showing Coulomb blockade diamonds with $2e$-periodicity in $V_\mathrm{g}$ for $V_\mathrm{sd}~<~\Delta$ and $1e$-periodicity for $V_\mathrm{sd}~>~\Delta$. \textbf{c}, Zero bias conductance, $G$ vs. $V_\mathrm{g}$ and temperature, $T$. 1e-periodic peaks above $T = 250-300$~mK emerge due to thermally excited quasi-particles. \textbf{d}, A single, continuous measurement of $G$ vs. $V_{\mathrm{g}}$ and $B_{\parallel}$, presented without corrections for, e.g., switching. 364 electrons were removed from the island. The 2e- to 1e-periodic transitions in the range $B_{\parallel}~=~0.15$~--~$0.25$~T occur as bound states move to zero energy. This can occur with \textbf{e}, equal spacing between $1e$ peaks, or \textbf{f}, $B_\parallel$-dependant spacing, depending on the nature of the bound states.\cite{AlbrechtNature16, VaitiekenasPRL18sag,ShenNatComm18} \textbf{g}, A parity transition to a state where only odd numbers of electrons are on the island is also possible.\cite{ShenNatComm18}}
\end{figure*}

\begin{figure}[t]
\includegraphics[width=8.5cm]{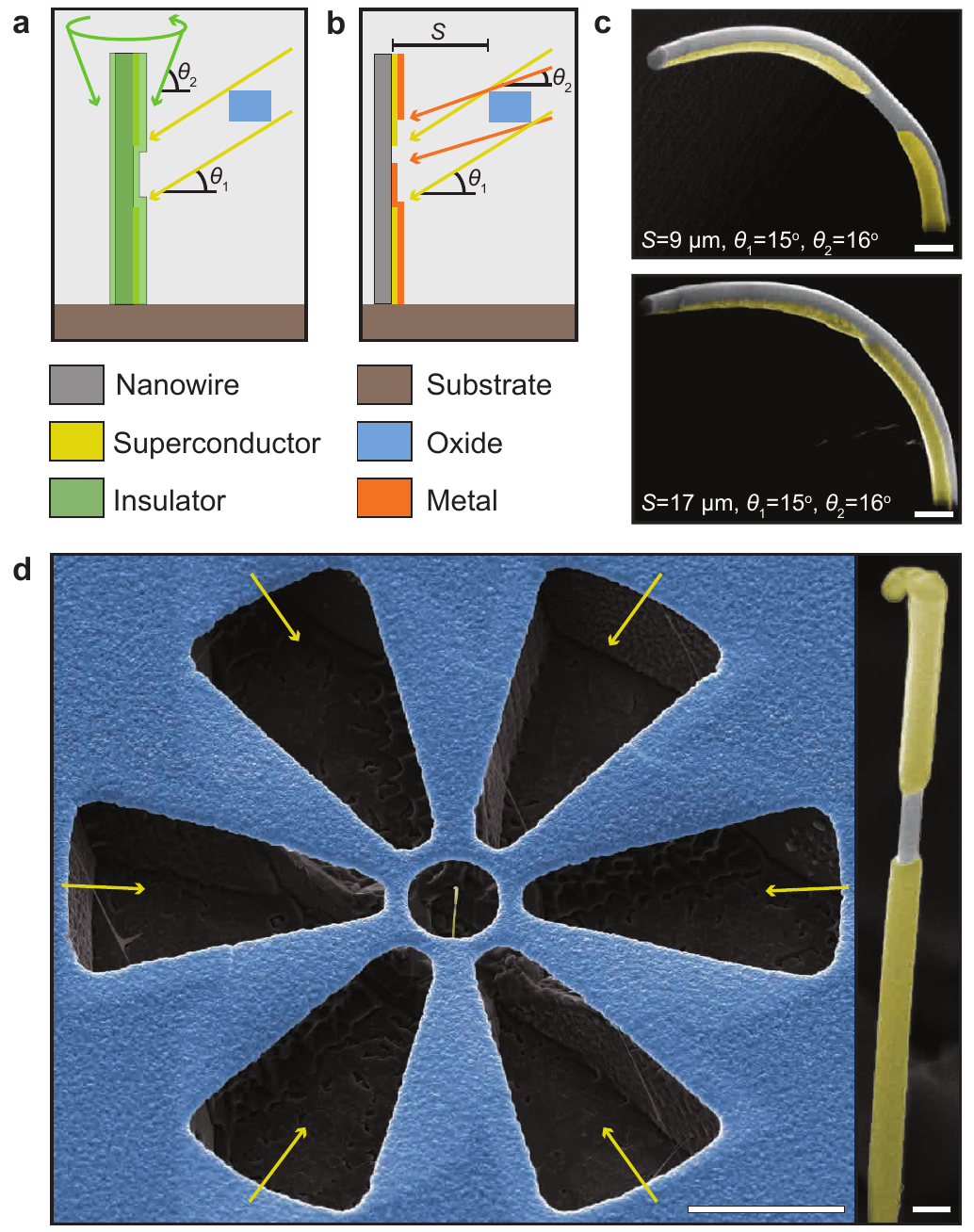}
\caption{\textbf{Advanced device geometries a}, Schematic of two-angle procedure for Josephson, or metal-NW-metal junctions protected by an \emph{in-situ} conformal insulator coating. \textbf{b}, Contacting the NW with two different materials is possible using evaporation of the desired materials from two different angles, $\theta_1$ and $\theta_2$. The pictured example constitutes an \emph{in-situ} formed tunnel spectroscopy device. \textbf{c}, Example of vanadium Josephson junctions realised by two-angle deposition, where junction length depends on $\theta_1$, $\theta_2$ and the NW-bridge separation, $S$. \textbf{d}, Demonstration of a full-shell extension of the shadow concept. Depositing from six different angles matching a six-fold symmetric bridge structure aligned with the facets of the NW, yields a full-shell JJ. Scale bars represent \textbf{c}, 200~nm, \textbf{d}, $10~\mu$m (left) and 200~nm (right).} 
\end{figure}

\section*{Stability of \emph{in-situ} Majorana island devices}
While tunnel spectroscopy devices are the essential tool for characterising induced superconductivity, the elemental building block in most topological quantum computing architectures are `Majorana islands',\cite{AasenPRX16, PluggeNJP17} which consist of a finite length hybrid segment retaining a charging energy $E_C$. Single\cite{AlbrechtNature16} and double\cite{ShermanNatNano16} island geometries with arbitrary superconductors are realised using a shadow epitaxy pattern of two or three bridges as shown in Figure\ 1f,g. The bridge separation(s) and widths determine the lengths of the hybrid segment(s) $(\propto E_C$) and the exposed NW segments, respectively. \textbf{Figure~4}a shows a typical Al-based shadow-patterned device, measured to enable comparison with conventional etched devices. The gate potentials $V_\mathrm{g}$ and $V_\mathrm{sg1,sg2}$ control the charge on the island and the tunnel coupling to the Ti/Au leads. \cite{AlbrechtNature16,HigginbothamNatPhys15,ShenNatComm18,ProutskiPRB19} For $E_C < \Delta$ and in the absence of quasi-particles or sub-gap states, charge is added in units of 2$e$ (Cooper pairs). Figure 4b, shows $G$ vs.\ $V_\mathrm{sd}$ and $V_\mathrm{g}$ exhibits $2e$-periodic Coulomb diamonds that correspond to a charging energy $2E_\mathrm{C}~\sim~130~\mu$eV (dashed lines). Quasi-particle 1$e$ periodic charging is seen for $|eV_{\mathrm{sd}}| > \Delta~\sim~180~\mu$eV (solid lines).\cite{AlbrechtNature16,HigginbothamNatPhys15} Figure~4c shows the temperature dependence of the low-bias $V_\mathrm{g}$-induced charging. The $2e$ periodic state persists up to $T~\sim 250-300$~mK, where a transition to $1e$ periodicity occurs due to thermal excitation of quasi-particles. This temperature is set by $\Delta$ and the volume of the superconductor and is comparable to that observed in etched devices.\cite{HigginbothamNatPhys15, VaitiekenasPRL18sag,ProutskiPRB19}

A reproducible and quiet electrostatic platform is crucial for charge-based topological devices, since environmental charge fluctuations and noise couple both to the island charge and tunnel couplings through the semiconductor, and thereby constitute a direct source of qubit decoherence.\cite{Knapp:2018,Li:2018} Furthermore, with increasing complexity of topological Majorana circuits, fine-tuning in a multi-dimensional space of cross-coupled parameters is required, and stability and reproducibility become important requirements. In conventional etched devices, the stable operation range is typically limited to $\lesssim 20$ consecutive charge states,\cite{AlbrechtNature16, HigginbothamNatPhys15, ShenNatComm18, VaitiekenasPRL18sag, ProutskiPRB19} as uncontrolled discrete charging of nearby impurity sites leads to random switches of the island charge/parity. The shadow device exhibits an increase of this range by at least an order of magnitude, consistent with a cleaner electrostatic environment due to the obviation of processing. This is demonstrated in Figure~4d, which shows the stable evolution of 182(364) consecutive 2$e$(1$e$) charge states accessed by continuously sweeping $V_\mathrm{g}$, and an interleaved, step-wise increase of $B_\parallel$. The island is superconducting in this field range, and the bifurcation of the spectrum at $B_\parallel \sim 200 \, \mathrm{mT}$ is caused by the appearance of a bound state below the gap, as previously analysed.\cite{AlbrechtNature16,VaitiekenasPRL18sag,ShenNatComm18} In addition to a topological transition, various trivial effects can lead to a $2e-1e$-transition in $B_\parallel$, each exhibiting distinct peak spacing and amplitude modulation.\cite{AlbrechtNature16,VaitiekenasPRL18sag,ShenNatComm18} For example, the zooms of three regions of panel Figure~4d, shown in Figures 4e-g, exhibit strikingly different behaviors. Figure~4e features $1e$-spacings independent of $B_\parallel$, in Figure~4f the peaks spacing and amplitude are modulated by $B_\parallel$,\cite{HansenPRB18} while a $2e$-$2e$ transition occurs with a parity change in the range of Figure\ 4h.\cite{ShenNatComm18} The behaviour shown in Figure~4f is consistent with a topological transition;\cite{AlbrechtNature16,ShenNatComm18,Vaitiekenasfullshell18,VaitiekenasPRL18sag,HansenPRB18,ShermanNatNano16} however, a detailed study is outside the scope of this work and will be presented elsewhere. The important point here is that the shadow platform greatly enhances the range of hysteresis-free, stable device operation, which in turn facilitates the simultaneous study of different behaviours. Measuring over this larger, stable range provides new information, including the identification of slowly varying features such as the $V_\mathrm{g}$-dependence of the $2e$-$1e$ transitions, which may be linked to gate-dependent $g$-factor of the bound state\cite{VaitiekenasPRL18gfactor} or related to weakly coupled bound-states localised in the leads.\cite{LeeNatNano14} The potential for increased understanding of device features and large range stability may thus enable reliable and rapid identification of the topological regime. In addition, stable and non-hysteretic navigation in parameter space is a prerequisite for employing automatic tuning and operation procedures that will be essential for operation of complex quantum devices in the future.\cite{LennonNPJQuantInf19} 

Just as the Majorana island is the basic component for topological architectures, gate-tunable Josephson junctions (JJs) constitute the elementary component of superconducting `gatemon'\cite{LarsenPRL15, LuthiPRL18} and Andreev qubit devices.\cite{vanWoerkomNatPhys17,LarocheNatComm19,TosiPRX19,HaysPRL18} Electrostatic stability is also critical here for qubit decoherence and tuning.\cite{LuthiPRL18} \emph{In-situ} JJ devices are grown using single-bridge shadows (Figure\ 1e), and low-temperature characterisation is presented in Supplementary Section 8, demonstrating quiet operation and gate-tunable critical currents $\sim 10$~nA, consistent with previous devices.\cite{LarsenPRL15, vanWoerkomNatPhys17,LarocheNatComm19}

\section*{Advanced architectures}
Having demonstrated the use of shadow epitaxy to realise the most important hybrid device geometries, we now discuss further extensions and the flexibility of the method. Firstly, \textbf{Figure~5}a illustrates how steep-angle depositions under rotation enable \textit{in-situ} conformal dielectric coatings that protect sensitive interfaces, contacts, and surfaces during subsequent device processing. A second extension, illustrated in Figure~5b, employs consecutive shadow depositions from different angles to realise a complete \emph{in-situ} junction with non-identical contact elements. Such double-angle-generated structures thus produce both lateral and axial hybrid devices incorporating, e.g., normal metal, superconducting and/or magnetic elements with pristine epitaxial interfaces. Double-angle deposition also enables JJ devices with arbitrarily short junction length, as demonstrated in Figure~5c for a vanadium based hybrid\cite{BjergfeltNanotech19} made by two depositions from different angles $\theta_1 = 16^{\circ}$, $\theta_2 = 17^{\circ}$. Conveniently, the semiconductor segment length $l_\mathrm{SE}$ depends not only the angles $\theta_1$ and $\theta_2$, but also the separation $S$ between bridge and NW. Increasing $S$ from $9~\mu$m to $17~\mu$m reduced $l_\mathrm{SE}$ from $\sim400$~nm to $\sim40$~nm. 

An important feature of our shadow epitaxy platform is the straightforward generalisation to full-shell geometries, achieved by radially copying the bridge design and depositing the coating from corresponding angles around the NW. Figure 5d shows an example full-shell JJ,\cite{Vaitiekenasfullshell18} and the other geometries from Figure~1 can be implemented following the same strategy. Deposition of different materials from the various directions further increases the possible functionality. Shadow epitaxy also applies to planar structures such as selective area grown nanostructures\cite{KrizekPRM18, VaitiekenasPRL18sag, SchuffelgenNatNano19} or vapour-liquid-solid NWs grown parallel to the substrate\cite{KrizekNL17} (Supplementary Section 9). Finally, we discuss the potential for large scale fabrication using vertical device structures in Supplementary Section 10. 

\section*{Conclusion}
\emph{In-situ}-grown superconductor/semiconductor hybrid devices form the backbone of electronic implementations of quantum information ranging from topological qubits, gatable transmon devices and Andreev quantum dots. The shadow epitaxy platform transcends previous restrictions on the possible material combinations, and obviates the most potentially damaging fabrication steps, thus providing clear enhancement of device quality and functionality. Beyond these applications, electrical contact quality, reproducibility and surface disorder play an pivotal role in nearly \emph{all} nanoscale devices\cite{delAlamoNat11, RielMRS14} and thus improving material quality \emph{via} shadow epitaxy holds the potential to strongly impact also conventional electronics, sensor-applications and optoelectronic devices.\cite{delAlamoNat11, RielMRS14}

\section*{Experimental Section}
\textbf{Fabrication of the shadow epitaxy platform.} The platform is based on InAs (111)B-oriented substrates, capped with a $100-150$~nm thick SiOx layer grown by plasma-enhanced chemical vapor deposition. The fabrication procedure is illustrated schematically in Supplementary Figure~1. The bridges were formed by two sequential wet etching steps using a single photolithographically defined resist mask (photoresist AZ1505), with the custom pattern exposed using a Heidelberg $\mu$PG501 LED writer. Etching the SiOx using buffered hydroflouric acid ($6\%$ in H$_2$O at room temperature) leaves strips of oxide in the desired bridge pattern, with typical length $L~\sim~10~\mu$m and widths in the range $W = 400$~nm -- $2~\mu$m. The resist was removed and the InAs was subsequently etched using a 37:23:12 solution of C$_6$H$_8$O$_7$(40\%):H$_3$PO$_4$(80\%):H$_2$O$_2$ with the SiOx pattern acting as an etch mask. The anisotropic etch preferentially exposes (111)B InAs crystal planes and thereby forms trenches in the substrate, with a (111)B surface at the trench bottom suitable for NW growth. An etch time of 5~mins at room temperature produced trences with depth $D \sim 7~\mu$m. The InAs etchant under-etches material from both sides of each SiOx strip and a void forms underneath the strips since their width $W \ll D$. The result is SiOx bridges suspended above the etched InAs surface (Figure~1a). The process is highly reproducible and robust on a wafer scale (Supplementary Figure~2). The potential to use the platform with antimonide, nitride, silicon, and other NW materials is discussed in Supplementary Section 1. Electron beam lithography (PMMA A4.5 and EL13 resist stack) and e-beam deposition of Au were used to define the catalyst particles for NW growth. The platform is naturally compatible with multiple catalyst particles per bridge, and methods for randomly defining catalyst position. Note that fabricating the bridges requires merely a single photolithography step; the complex structures are obtained \emph{via} relatively simple processing.

\textbf{Nanowire hybrid growth.}
InAs NWs were grown \emph{via} the Au catalyst-assisted vapour-liquid-solid mechanism in a solid-source Varian GEN-II MBE system following a two-step protocol. In the first step, NWs were grown vertically along the (111)B direction, using As\textsubscript{4} cracker temperature $500^\circ$C, substrate thermocouple temperature of $447^\circ$C and V/III flux ratio $\sim~10$. A growth time of 80~mins resulted in $5-8~\mu$m-long NWs. The second step involved growth at reduced substrate thermocouple temperature ($350^\circ$C) and increased As cracker temperature ($800^\circ$C) for 10 mins. These conditions promote radial overgrowth and flatter NW facets, optimal for subsequent superconductor deposition.

After InAs growth, the substrate was transferred under ultra-high vacuum to a metal evaporation chamber, with a freely rotatable sample holder that can be cooled to a thermocouple temperature of approx. $-150^\circ$C using liquid nitrogen. The low temperature is used to decrease the size and pitch between the initial critical clusters, which promotes a continuous and flat film morphology.\cite{KrogstrupNatMat15} It also suppresses the potential for parasitic material to settle outside the shadow-mask defined regions, which may occur in particular for Al depositions at higher temperatures. The superconductor layers (Al, Ta, Nb, V) were deposited using e-beam evaporation at a fixed angle to align with the shadow mask design. The NWs grow with type-I, $\{1\overline{1}00\}$ facets and the shadow mask design is oriented such that the superconductor deposition coats either 2 or 3 of the NW facets, as desired.

\textbf{Electron microscopy.}
SEM characterisation of the substrates, as-grown NWs and finished devices was carried out in a JEOL 7800F using acceleration voltage $V_{acc} = 5-10$~kV. For TEM characterisation, NWs were transferred from the growth substrate to a carbon membrane grid using a micromanipulator under an optical microscope. TEM micrographs were obtained using either a Philips CM20 (Figure 2, center panels) with $V_{acc} = 200$~kV, a FEI Titan Analytical 80-300ST featuring a monochromator and $V_{acc} = 300$~kV (Figure 2a, right panel) or a Phillips 3000F with $V_{acc} = 300$~kV (Figures. 2b,c, right panels). NWs were oriented with the beam parallel to an InAs facet ($\left\langle2\bar{1}\bar{1}0\right\rangle$ zone-axis).

\textbf{Device fabrication and measurement.} NWs were transferred from the growth substrate to pre-patterned device substrates using a micromanipulator under an optical microscope. Ti/Au contacts and side gates were defined simultaneously using electron beam lithography and e-beam deposition. Ar$^+$-ion milling was used immediately prior to contact deposition to remove native InAs oxide and ensure ohmic contact to the NW. Electron transport measurements were conducted in an Oxford Triton dilution refrigerator with a base temperature $\sim 20$~mK and a 6-1-1~T vector magnet. Standard low frequency ($\sim 200$~Hz) lock-in techniques were used to measure differential conductance. For the island device in Figure~4, the cross-coupling of $V_{\mathrm{g}}$ to the tunnel barrier segments was compensated by sweeping $V_{\mathrm{sg1,sg2}}$ simultaneously with $V_{\mathrm{g}}$ using experimentally determined proportionality factors. Note the small offsets in Figures~4d-g between measurements at successive $V_\mathrm{g}$ vanish when measuring within a smaller gate range, as in Figures~4b,c. All electrical data throughout this work are presented `raw', without any corrections for switching events.


\begin{acknowledgements}
This work was funded by Innovation Fund Denmark's Quantum Innovation Center Qubiz, the Carlsberg Foundation, the Danish National Research Foundation, the Villum Foundation, European Union’s Horizon 2020 research and innovation programme FETOpen grant no. 828948 (AndQC) and QuantERA project no. 127900 (SuperTOP). We thank Saulius Vaitiekenas, Asbjørn Drachmann, Juan Carlos Estrada Saldaña, Kasper Grove Rasmussen, Dags Olsteins, Mikelis Marnauza, Lukas Stampfer, Joachim Sestoft, David van Zanten, Peter Krogstrup and I-Ju Chen for many helpful discussions. We also thank Shivendra Upadhyay, Claus B. Sørensen, Ajuan Cui and Thomas Pedersen (DTU Nanolab) for technical assistance and Charline Kirchert and Bilal Kousar for assistance with fabrication.
\end{acknowledgements}



\section*{Supporting Information}
\textbf{Supporting Information} is available at \url{https://sid.erda.dk/share_redirect/BxJJoUWo0N}. Raw electrical data and images for all figures are available from \url{https://doi.org/10.17894/ucph.3a22ad55-37fb-4495-b21d-4294e1e57fcc}

\bibliography{Bridges}
\bibliographystyle{advmatbib}

\end{document}